\documentstyle[aps]{revtex}
\begin{document}
\draft
\title{
Kolmogorov Scaling for the $\epsilon$-entropy 
in a Forced Turbulence Simulation 
\cite{N}}
\author{
W. Sakikawa and O. Narikiyo}
\address{
Department of Physics, 
Kyushu University, 
Fukuoka 810-8560, 
Japan}
\date{
August, 2002}
\maketitle
%----------------------------------------------------------------
\begin{abstract}
We have shown that 
the $\epsilon$-entropy determined from the time-series 
of the velocity fluctuation in a forced turbulence 
simulated by the lattice Boltzmann method 
obeys the Kolmogorov scaling. 
\end{abstract}
%----------------------------------------------------------------
\vskip 18pt
  Turbulence is one of the typical phenomena 
with multiscale motions. 
  Each phenomena of a scale strongly couples with 
all the other scales of turbulent motion. 
  In order to analyze such a system 
a scale-dependent entropy, the so-called $\epsilon$-entropy, 
works well.\cite{Vul} 
  For example, the time series of the velocity fluctuation in turbulence 
leads to a non-trivial scaling relation of the $\epsilon$-entropy, 
$ h(\epsilon) \propto \epsilon^{-3} $, 
expected from the Kolmogorov scaling.\cite{Vul} 
  The existing experimental data are consistent with this scaling.\cite{Vul} 
  In this Short Note we try to show the consistency of the scaling 
with the simulation on the basis of the lattice Boltzmann method 
which is one of the easiest ways to simulate turbulence. 

  The distribution function $f_i({\vec r},t)$ 
of the particle with the velocity ${\vec c}_i$ 
at position ${\vec r}$ and time $t$ 
obeys the lattice Boltzmann equation\cite{S,WG} 
%----------------------------------------------------------------
\begin{equation}
f_i({\vec r}+{\vec c}_i,t+1) - f_i({\vec r},t)
= - \omega [ f_i({\vec r},t) - f_i^0({\vec r},t) ],
\end{equation}
%----------------------------------------------------------------
where we have adopted the Bhatnagar-Gross-Krook (BGK) model 
for the collision term 
and the time step has been chosen as unity. 
  The position ${\vec r}$ of the particle 
is restricted on the cubic lattice 
and the lattice spacing is chosen as unity. 
  The local equilibrium distribution $f_i^0({\vec r},t)$ is assumed 
to be given as 
%----------------------------------------------------------------
\begin{equation}
f_i^0({\vec r},t)
= w_i \rho [ 1 + 3 ({\vec c}_i\cdot{\vec u}) 
               + {9 \over 2} ({\vec c}_i\cdot{\vec u})^2 
               - {3 \over 2} ({\vec u}\cdot{\vec u}) ].
\end{equation}
%----------------------------------------------------------------
  The density $\rho$ and the velocity ${\vec u}$ of the fluid 
are given by the sum of the contributions of the particles as 
$\rho = \sum_i f_i({\vec r},t)$ and 
$\rho {\vec u} = \sum_i f_i({\vec r},t) {\vec c}_i$, respectively. 
  Using the multiscale analysis 
the Navier-Stokes equation, 
which is supposed to be able to describe turbulence, 
is derived from the present BGK lattice Boltzmann equation 
and the corresponding viscosity $\nu = (1/\omega - 1/2)/3$. 

  In the following 
we use a 15-velocity model and ${\vec c}_i$ is chosen as 
${\vec c}_1 = (0,0,0)$, 
${\vec c}_i = (\pm1,0,0),(0,\pm1,0),(0,0,\pm1)$ for $i=2,3,\cdot\cdot\cdot,7$ 
and 
${\vec c}_i = (\pm1,\pm1,\pm1)$ for $i=8,9,\cdot\cdot\cdot,15$. 
  Then the weight factor $w_i$ for the local equilibrium distribution 
is determined as 
$w_1 = 2/9$, 
$w_i = 1/9$ for $i=2,3,\cdot\cdot\cdot,7$ and 
$w_i = 1/72$ for $i=8,9,\cdot\cdot\cdot,15$. 
  The relaxation frequency $\omega$ is chosen as $\omega=1.94$. 

  In order to simulate forced turbulence 
we add the forcing term ${\vec g}_i\cdot{\vec F}({\vec r},t)$ 
due to an applied body force ${\vec F}({\vec r},t)$ 
to the right hand side of eq.\ (1). 
  In order to reproduce the Navier-Stokes equation 
we should choose as ${\vec g}_i = {\vec c}_i/10$ 
for the present 15-velocity model.\cite{WG} 
  For simplicity we apply a solenoidal force in $y$-direction,\cite{S} 
${\vec F}=(0,F_y(z,t),0)$, and $F_y(z,t)$ is a Gaussian white noise 
whose variance $\sigma$ is a function of the lattice coordinate 
in $z$-direction, 
$\sigma = |0.01 \times \sin(2\pi k_f z / L)|$, 
where $L$ is the linear dimension of the simulation region 
of cubic box and $L=100$ ($1 \leq x,y,z \leq 100$). 
  We have adopted the periodic boundary condition. 
  Although we have chosen as $k_f = 4$, 
the results of the simulation is insensitive to the choice. 

  In the initial state of the simulation 
the fluid density is uniform, $\rho = 1$, and 
the spatial distribution of the fluid velocity ${\vec u}$ is Gaussian 
with the variance $\sigma = \sqrt{0.0075}$. 

  The Reynolds number $R$ for our simulation\cite{S,WG} is 
estimated as $R = \sqrt{ \langle {\vec u}^2 \rangle } L / 2\pi\nu$ 
and time-dependent as shown in Fig.\ 1. 
  Here $\langle \cdot\cdot\cdot \rangle$ represents the spatial average. 
  In our simulation 
we can realize a turbulent state for relatively small Reynolds number, 
since we add random force. 

  We focus our attention to the time series of 
the $y$-component $u_y({\vec r},t)$ of 
the fluid velocity ${\vec u}({\vec r},t)$ 
as shown in Fig.\ 2, 
since our system is anisotropic due to the forcing. 
  This turbulent signal is characterized by the exit-time 
$\tau(\epsilon;{\vec r})$ at position ${\vec r}$. 
  At the exit-time $t=\tau(\epsilon;{\vec r})$ 
the first exit satisfying the condition, 
$|u_y({\vec r},t_0+t) - u_y({\vec r},t_0)| > \epsilon/2$, 
occurs. 
  The time $t$ is measured from $t_0$ and $t_0=1000$ in our simulation 
regarding the states at the first 1000 time-steps as transitional. 

  The mean of the exit-time, 
$\tau(\epsilon) \equiv \langle \tau(\epsilon;{\vec r}) \rangle$, 
is related to the $\epsilon$-entropy, $h(\epsilon)$, as\cite{Vul} 
%----------------------------------------------------------------
\begin{equation}
h(\epsilon) \propto {1 \over \tau(\epsilon)}. 
\end{equation}
%----------------------------------------------------------------
  In the region where Kolmogorov's 4/5 law holds\cite{Vul,F} 
the velocity difference between two points behaves as 
$\langle|{\vec u}({\vec r}+{\vec R},t_0)-{\vec u}({\vec r},t_0)|\rangle 
     \propto |{\vec R}|^{1/3}$. 
  Under the assumption of Taylor's hypothesis,\cite{Vul,F} 
$[{\vec u}({\vec r},t_0+t)-{\vec u}({\vec r},t_0)] 
     \sim [{\vec u}({\vec r}+{\vec R},t_0)-{\vec u}({\vec r},t_0)]$, 
with $|{\vec R}|=Ut$, 
where $U$ is the mean velocity for large scale sweeping, 
we obtain 
%----------------------------------------------------------------
\begin{equation}
\langle|{\vec u}({\vec r},t_0+t)-{\vec u}({\vec r},t_0)|\rangle 
     \propto t^{1/3},
\end{equation}
%----------------------------------------------------------------
as a mean-field description. 
  Thus the definition of the exit-time leads to 
$\tau(\epsilon)^{1/3} \propto \epsilon$. 
  From eq.\ (3) we can conclude 
that $h(\epsilon) \propto \epsilon^{-3}$ 
in the above mean-field sense.\cite{Vul} 

  In accordance with the above scaling analysis 
our simulation data for $1/\tau(\epsilon)$ shown in Fig.\ 3 
have the scaling region for $\epsilon > \epsilon_0$ 
where $\epsilon_0 \sim 0.04$. 
  Thus by simulation we have established 
that $h(\epsilon) \propto \epsilon^{-3}$. 
  For $\epsilon < \epsilon_0$, namely for short-time events, 
we can not expect the scaling, 
since the mean-field assumption for large scale sweeping 
is valid for the average of many events and 
not applicable to a few events at short time-interval. 
  These behaviors are consistent with the experimantal data.\cite{Vul} 
  For $\epsilon \ll \epsilon_0$ 
the exit in the simulation occurs within the time step. 

  In summary 
we have shown that the velocity fluctuation in a forced turbulence 
simulated by the lattice Boltzmann method 
leads to the $\epsilon$-entropy scaling, 
$ h(\epsilon) \propto \epsilon^{-3} $, 
expected from the Kolmogorov scaling. 
  Although the scaling relation is a robust result 
of the mean-field description, 
the intermittent aspects of turbulence 
are beyond the scope of the present study. 

  This work was supported in part 
by a Grand-in-Aid for Scientific Research 
from the Ministry of Education, Culture, Sports, Science 
and Technology of Japan. 

%----------------------------------------------------------------

\vskip 15pt
%----------------------------------------------------------------
\begin{figure}
\caption{
The Reynolds number as a function of time. }
\label{fig:1}
\end{figure}

\vskip -15pt

\begin{figure}
\caption{
The $y$-component of the fluid velocity $u_y$ 
as a function of time at ${\vec r}=(50,50,50)$. }
\label{fig:2}
\end{figure}

\vskip -15pt

\begin{figure}
\caption{
The inverse of the mean exit-time $\tau(\epsilon)$. 
Here the dots and straight line represent the simulation data 
and $\epsilon^{-3}$ law, respectively. }
\label{fig:3}
\end{figure}
%----------------------------------------------------------------
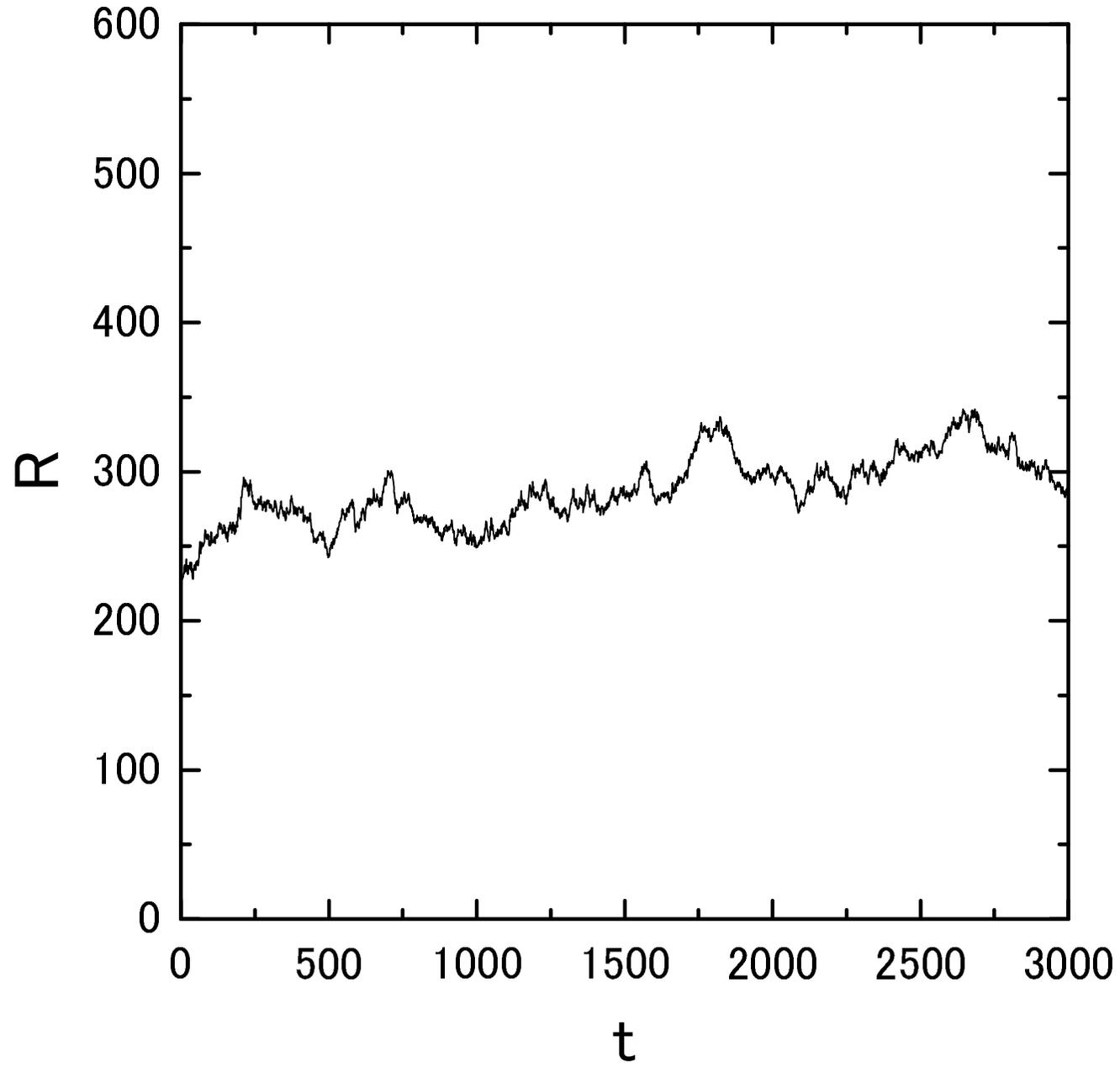
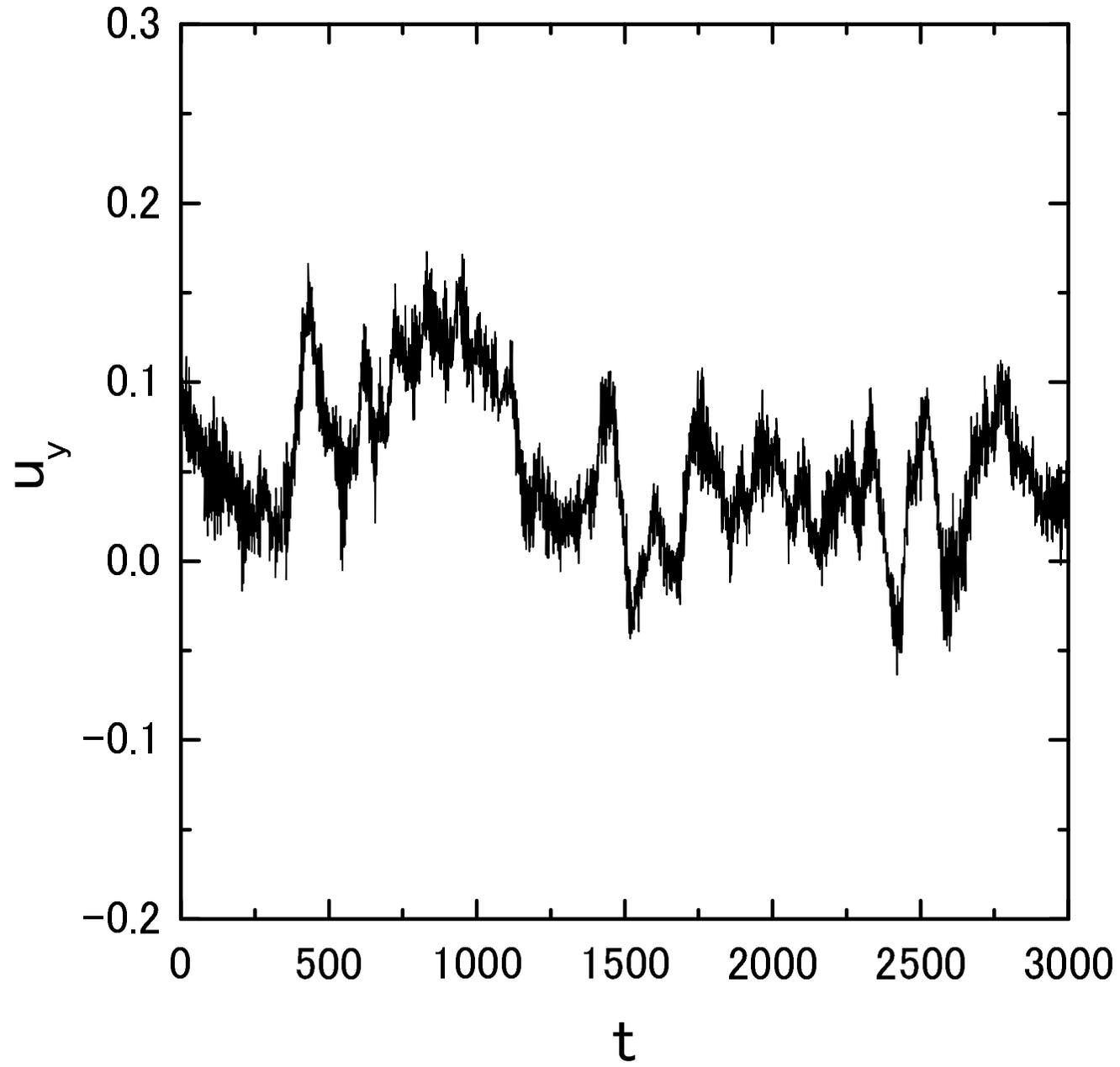
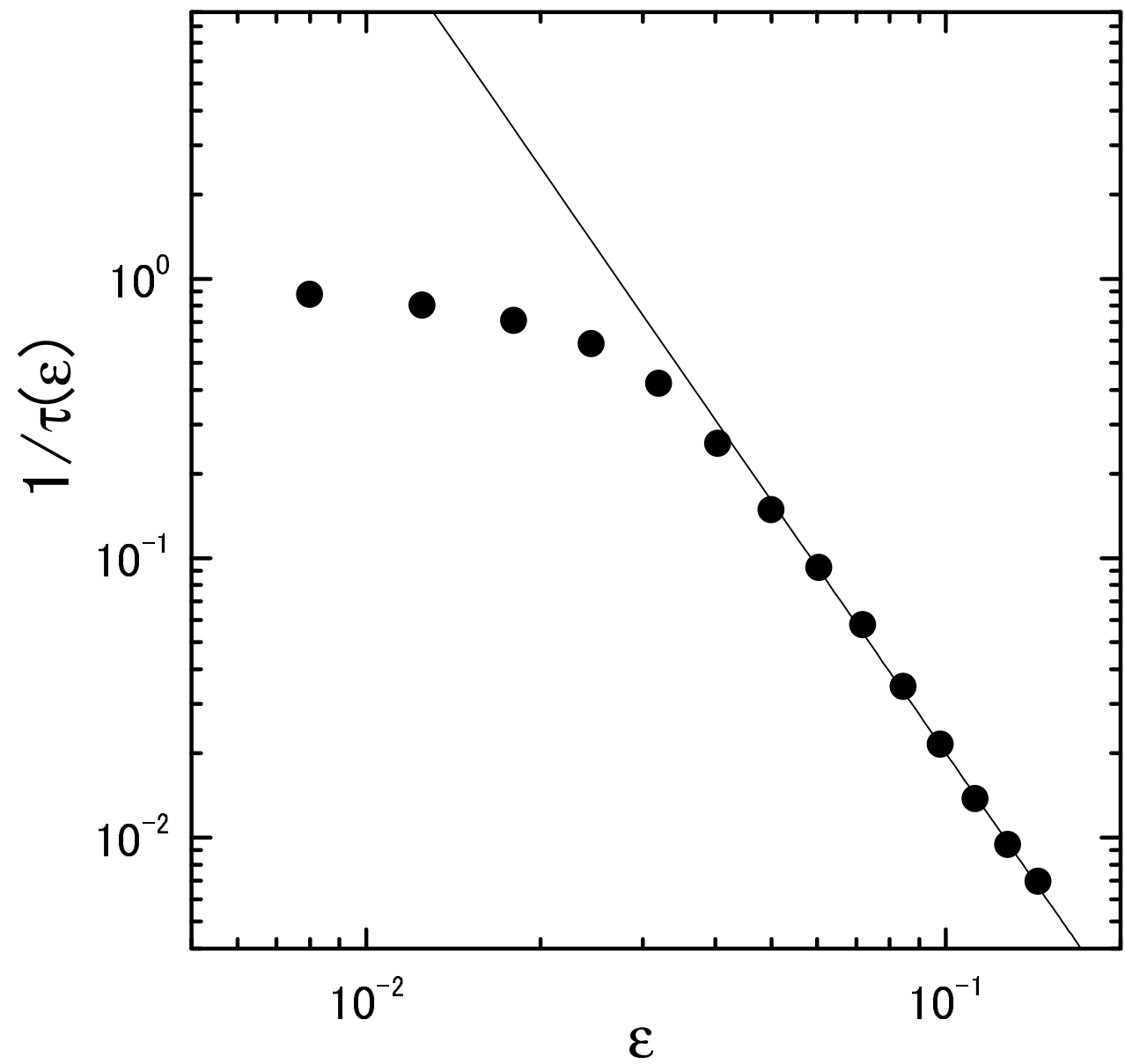
\end{document}